# Magnetic ordering and ferroelectricity in multiferroic $2H$-AgFeO$_2$: Comparison between hexagonal and rhombohedral polytypes

Noriki Terada,[1,*] Dmitry D. Khalyavin,[2] Pascal Manuel,[2] Yoshihiro Tsujimoto,[1] and Alexei A. Belik[3,†]

[1]*National Institute for Materials Science, Sengen 1-2-1, Tsukuba, Ibaraki 305-0047, Japan*
[2]*ISIS facility, STFC Rutherford Appleton Laboratory, Chilton, Didcot, Oxfordshire, OX11 0QX, United Kingdom*
[3]*International Center for Materials Nanoarchitectonics (WPI-MANA), National Institute for Materials Science,
1-1 Namiki, Tsukuba, Ibaraki 305-0044, Japan*

Magnetic and dielectric properties of the hexagonal triangular lattice antiferromagnet $2H$-AgFeO$_2$ have been studied by neutron diffraction, magnetic susceptibility, specific heat, pyroelectric current, and dielectric constant measurements. The ferroelectric polarization, $P \simeq 5$ $\mu$C/m$^2$, has been found to appear below 11 K due to a polar nature of the magnetic ground state of the system. In the temperature range of 11 K $\leqslant T \leqslant$ 18 K, an incommensurate spin density wave (ICM1) with the nonpolar magnetic point group $mmm1'$ and the $\boldsymbol{k}_1 = (0, q_b^1, 0; q_b^1 = 0.390$–$0.405)$ propagation vector takes place. Below 14 K, a proper screw ordering (ICM2) and $\boldsymbol{k}_2 = (0, q_b^2, 0; q_b^2 = 0.385$–$0.396)$ appears as a minor phase which coexists with ICM1 and the ground state down to the lowest measured temperature 5.5 K. No ferroelectric polarization associated with the ICM2 phase was observed in agreement with its nonpolar point group $2221'$. Finally, a spiral order with cycloid and proper screw components (ICM3), and $\boldsymbol{k}_3 = (q_a^3, q_b^3, 0; q_a^3 = 0.0467, q_b^3 = 0.349)$ emerges below 11 K as the ground state of the system. Based on the deduced magnetic point group $21'$, we conclude that the ferroelectric polarization in ICM3 is parallel to the $c$ axis and is caused by the inverse Dzyloshinskii-Moriya effect with $\boldsymbol{p}_1 \propto \boldsymbol{r}_{ij} \times (\boldsymbol{S}_i \times \boldsymbol{S}_j)$. Unlike the rhombohedral $3R$-AgFeO$_2$ polytype, the additional contribution to the macroscopic polarization $\boldsymbol{p}_2 \propto \boldsymbol{S}_i \times \boldsymbol{S}_j$ is not allowed in the present case due to the symmetry constraints imposed by the hexagonal lattice of $2H$-AgFeO$_2$.

## I. INTRODUCTION

Magnetoelectric mutliferroic materials with (anti)ferromagnetic and ferroelectric properties have attracted much attention, after the discovery of a magnetic-field-controlled ferroelectric polarization in TbMnO$_3$ [1]. In some multiferroics, noncollinear spin arrangements strongly correlate with the emergence of ferroelectric polarization [2,3]. The inverse Dzyaloshinskii-Moriya (DM) mechanism, expressed by $\boldsymbol{p} \propto \boldsymbol{r}_{ij} \times (\boldsymbol{S}_i \times \boldsymbol{S}_j)$ [4–6], has successfully explained the appearance of ferroelectric polarization for cycloid spin orderings in orthorhombic perovskites [2,3]. However, this relation is not applicable for proper screw spin structures, observed in many multiferroics [7–12], because the propagation vector is parallel to the cross product of neighboring spins $\boldsymbol{r}_{ij} || \boldsymbol{S}_i \times \boldsymbol{S}_j$. Kaplan and Mahanti have extended the inverse DM mechanism, which can explain the emergence of polarization induced by the proper screw ordering [13]. Johnson *et al.* have also explained the ferroelectricity associated with proper screw order by a coupling between the spin chirality, $\boldsymbol{r}_{ij} \cdot (\boldsymbol{S}_i \times \boldsymbol{S}_j)$, and ferroaxial crystal rotation, based on the inverse DM effect [8–10]. In these newly developed theories, the crystal symmetry in the paramagnetic phase of the multiferroic compounds is the crucial ingredient for understanding the coupling between noncollinear spin ordering and ferroelectric polarization.

Copper and silver delafossite compounds, $AB$O$_2$ ($A$ = Cu, Ag, $B$ = Cr, Fe, Ni, etc.) with the rhombohedral $R\bar{3}m$ space group [Fig. 1(b)], offer a rich playground for studying the multiferroic couplings [11,12,14–22]. The type of noncollinear spin order in delafossites, cycloid or proper screw, depends on both nonmagnetic $A$-site and magnetic $B$-site cations due to a strong spin frustration imposed by the triangular lattice illustrated in Fig. 1(b) [22]. Recently, some of us have reported the emergence of ferroelectric polarization associated with the cycloid magnetic structure in the rhombohedral delafossite AgFeO$_2$ ($3R$-AgFeO$_2$) [21]. In these systems, two magnetic phase transitions occur at 15 and 9 K [21,25]. While the magnetic ordering for 9 K $\leqslant T \leqslant$ 15 K is a collinear spin density wave (SDW), it turns into a cycloid ordering with the propagation vector $\boldsymbol{k} = (\bar{\frac{1}{2}}, q, \frac{1}{2})$ below 9 K, concomitant with the appearance of ferroelectric polarization $P \simeq 300$ $\mu$C/m$^2$ [21]. The ferroelectric polarization has been concluded to be driven by the inverse DM effect with two orthogonal components, $\boldsymbol{p}_1 \propto \boldsymbol{r}_{ij} \times (\boldsymbol{S}_i \times \boldsymbol{S}_j)$ and $\boldsymbol{p}_2 \propto \boldsymbol{S}_i \times \boldsymbol{S}_j$, based on the symmetry consideration and the extended theory [13].

The hexagonal polytype $2H$-AgFeO$_2$ also has a triangular-based layered structure [drawn in Fig. 1(a)] similar to the rhombohedral one. However, the two polytypes crystallize in different space groups, $R\bar{3}m$ in $3R$-AgFeO$_2$ and $P6_3/mmc$ in $2H$-AgFeO$_2$. When the incommensurate order occurs in $3R$-AgFeO$_2$ at low temperature, it breaks three-fold rotational symmetry and the space group which leaves the magnetic propagation vector $\boldsymbol{k}$, unchanged or changes to $-\boldsymbol{k}$ (extended wave vector group) lowers into $C2/m$. It is convenient to use the extended wave vector group to discuss the symmetry-allowed components of the spin-induced polarization and the associated mechanisms [21]. In particular, the monoclinic $C2/m$ symmetry allows the $\boldsymbol{p}_2 \propto \boldsymbol{S}_i \times \boldsymbol{S}_j$ component of the polarization induced by the ferroaxial mechanism or the other through the inverse DM effect. On the other hand, if a similar incommensurate ordering breaks the six- and three-fold symmetry in the hexagonal $2H$-AgFeO$_2$ with $P6_3/mmc$, the extended wave vector group would be the orthorhombic $Ccmm$ that does not allow $p_2$ and the ferroaxial coupling. Therefore, a comparative study of the spin ordering and ferroelectricity in $3R$-AgFeO$_2$ and $2H$-AgFeO$_2$ offers a great opportunity to comprehensively explore the extended inverse DM mechanism [13]. The magnetic and dielectric properties of $2H$-AgFeO$_2$ polytype, however, have not been investigated so far.

*TERADA.Noriki@nims.go.jp
†Alexei.Belik@nims.go.jp

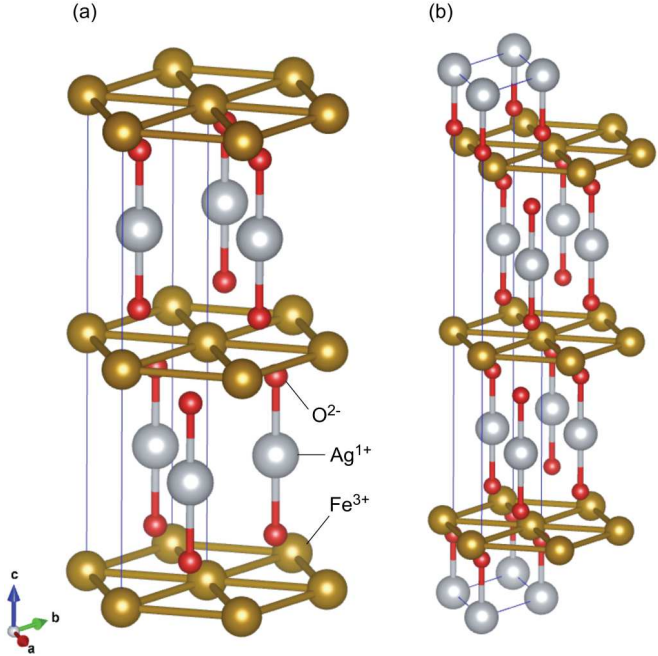

FIG. 1. (Color online) Crystal structures of (a) hexagonal $2H$-AgFeO$_2$ [$P6_3/mmc$; $a = 3.039$ Å, $c = 12.395$ Å; Ag(1/3,2/3,0.25), Fe(0,0,0), O(1/3,2/3,0.0833) [23] and (b) rhombohedral $3R$-AgFeO$_2$ [$R\bar{3}m$; $a = 3.0391$ Å, $c = 18.590$ Å; Ag(0,0,0), Fe(0,0,0.5), O(0,0,0.1112) [24].

In the present study, we have explored the magnetic and dielectric properties of the hexagonal $2H$-AgFeO$_2$ polytype by means of neutron diffraction, magnetic susceptibility, specific heat, pyroelectric current, and dielectric constant measurements. The obtained results are compared to the rhombohydral $3H$-AgFeO$_2$ counterpart and the difference in the dielectric properties is discussed based on the complex noncollinear magnetic structures taking place in these systems.

## II. EXPERIMENTAL DETAILS

Powder specimens of $2H$-AgFeO$_2$ were prepared from a stoichiometric mixture of $\alpha$-Fe$_2$O$_3$ (99.999%) and Ag$_2$O (99.99%). The mixture was placed in Au capsules and treated at 6 GPa in a belt-type high-pressure apparatus at 1073 K for 40 min (heating time to the desired temperature was 5 min). After the heat treatment, the samples were quenched down to room temperature, and the pressure was slowly released. The synthesis at 3 GPa and 1073 K for 40 min produced $3R$-AgFeO$_2$. Powder x-ray diffraction demonstrated that they are almost single phase with a tiny amount of $\alpha$-Fe$_2$O$_3$ impurity phase. For magnetic susceptibility measurements, we used the Magnetic Properties Measurement System (MPMS) manufactured by Quantum Design (QD). For specific heat measurements, we used QD's Physical Properties Measurement System (PPMS). The dielectric constant and pyroelectric current measurements were done by using an Agilent E4980A LCR meter and Keithley 6517B electrometer. The dielectric properties were measured using hardened pellets (5.9 mm$^2$ covered with silver paste $\times 1.0$-mm thickness) of the polycrystalline $2H$-AgFeO$_2$ material. The neutron powder diffraction measurements were carried out on the cold neutron time-of-flight diffractometer WISH [26] at ISIS Facility, UK. Crystal and magnetic structure refinements were performed using the FULLPROF program [27].

## III. EXPERIMENTAL RESULTS

### A. Bulk properties

#### 1. Magnetic susceptibility and specific heat

Figure 2 shows a temperature dependence of the magnetic susceptibility and specific heat of $2H$-AgFeO$_2$, which demonstrates the presence of two magnetic phase transitions. A broad peak anomaly was observed at $T = 18$ K ($\equiv T_{N1}$) in the magnetic susceptibility, while the specific heat also shows a local minimum in $d^2C/dt^2$ at $T_{N1}$ [inset of Fig. 2(b)]. These results infer that a magnetic phase transition from paramagnetic to a magnetically ordered phase occurs at $T_{N1}$. We also observed another anomaly at the lower temperature, 11 K ($\equiv T_{N2}$); a sharp decrease in magnetic susceptibility and broad peak in the specific heat. This anomaly corresponds to the second magnetic phase transition.

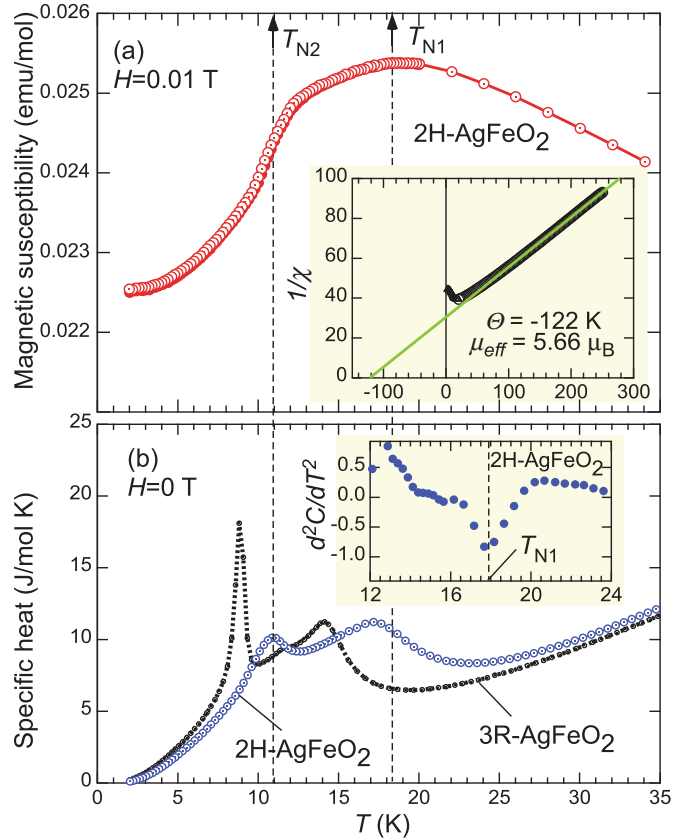

FIG. 2. (Color online) Temperature dependence of (a) magnetic susceptibility and (b) specific heat of $2H$-AgFeO$_2$. The inset in (a) shows the inverse magnetic susceptibility with Curie-Weiss law fitting (solid line). Closed symbols in (b) denote the data for $3R$-AgFeO$_2$. Dotted lines indicate temperatures, where the magnetic phase transitions occur in $2H$-AgFeO$_2$.

In the previous work for the rhombohedral $3R$-AgFeO$_2$ polytype, two magnetic phase transitions at 15 and 9 K were reported [21]. As is also shown in Fig. 2(b), the width of the anomaly at the higher transition temperatures is similar between the two systems, while the peak anomaly at the lower phase transition for $3R$-AgFeO$_2$ is much sharper than that in $2H$-AgFeO$_2$. The results suggest that the entropy change at the second phase transition is different for these two systems.

As shown in the inset of Fig. 2(a), the inverse magnetic susceptibility in the temperature range of 150 K $\leqslant T \leqslant$ 300 K can be described by the Curie-Weiss law. The refined Weiss temperature and the effective magnetic moment in $2H$-AgFeO$_2$ are $\Theta = -122$ K and $\mu_{eff} = 5.66\ \mu_B$, respectively. The $\mu_{eff}$ is slightly lower than the spin-only value of $S = 5/2$ expected for Fe$^{3+}$, $\mu_{eff} = 5.92\ \mu_B$. The $\Theta$ in $2H$-AgFeO$_2$ is higher than that in $3R$-AgFeO$_2$, $\Theta = -140$ K [25], suggesting that the average antiferromagnetic exchange interactions in the former are slightly weaker than in the latter.

### 2. Dielectric property

Ferroelectric polarization was observed below $T = 11$ K ($T_{N2}$) in the $2H$-AgFeO$_2$ polytype, as shown in Fig. 3(a). As follows from the inset of Fig. 3(a), the poling electric field dependence of the polarization becomes almost flat at $P = 5\ \mu C/m^2$ in the polycrystalline sample, which implies the intrinsic value of the polarization to be $P_{intrinsic} \simeq 10\ \mu C/m^2$. The polarization in $2H$-AgFeO$_2$ is two orders of magnitude smaller than that in $3R$-AgFeO$_2$, $P \simeq 300\mu C/m^2$ [21]. A small hump was observed at $T \simeq 9$ K as indicated by an arrow in Fig. 3(a), which might be caused by a tiny amount of impurity in the phase of $3R$-AgFeO$_2$ with the large ferroelectricity below $T = 9$ K [21]. As shown in Fig. 3(b), the dielectric constant shows a kink anomaly at 11 K for all frequencies measured, which is in agreement with the ferroelectric phase transition.

## B. Neutron diffraction

### 1. Temperature dependence of magnetic reflections

A temperature dependence of the neutron diffraction profiles for $2H$-AgFeO$_2$ is shown in Fig. 4. With decreasing temperature from the paramagnetic phase, a set of magnetic Bragg reflections appears below $T_{N1} = 18$ K. Note here that we observed a peak broadening for several nuclear reflections with indexes including either hexagonal $h$ or $k$ components, such as 110 and 200, below $T_{N1}$, collected with the high-resolution backscattering detector banks of the WISH diffractometer. The broadening indicates losing the six- and three-fold rotational symmetry elements of $P6_3/mmc$, and implies symmetry lowering down to at least the maximal nonisomorphic subgroup $Ccmm$ (No. 63), which takes into account only the coupling of the magnetic order parameter to the macroscopic strains. Note that this orthorhombic subgroup is actually the extended group of the magnetic propagation vector discussed below. Hereafter, we use the orthorhombic setting $\boldsymbol{a} = -\boldsymbol{a}_{hexa} + \boldsymbol{b}_{hexa}$, $\boldsymbol{b} = \boldsymbol{a}_{hexa} + \boldsymbol{b}_{hexa}$ and $\boldsymbol{c} = -\boldsymbol{c}_{hexa}$, unless specified. The observed magnetic reflections can be assigned to the incommensurate propagation vector $\bm{k}_1 = (0, q_b^1, 0)$ with $q_b^1 \simeq 0.39$–$0.40$ in the orthorhombic notation [corresponding to $(q_b^1/2, q_b^1/2, 0)$ in the hexagonal cell]. (Hereafter, we call the magnetic order with $\bm{k}_1$ as ICM1.) As shown in Fig. 5(b), the component of the $\bm{k}_1$ vector significantly depends on temperature and shows a thermal hysteresis, which is similar to that seen in the intermediate temperature phase of $3R$-AgFeO$_2$ [21]. The spin correlation length, derived from the width of the magnetic peaks, depends on temperature and is in the ~400–1000 Å limit [Fig. 5(c)].

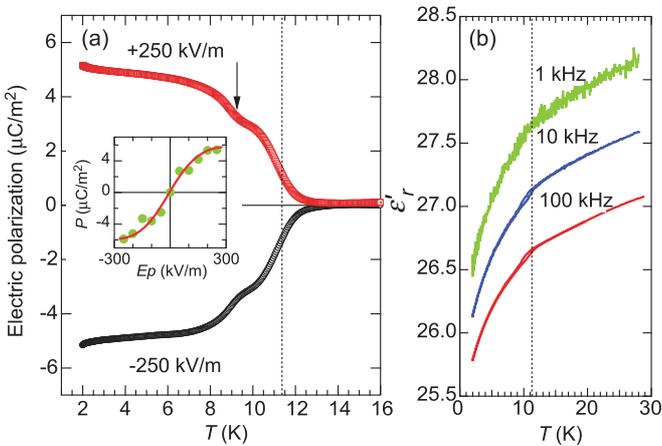

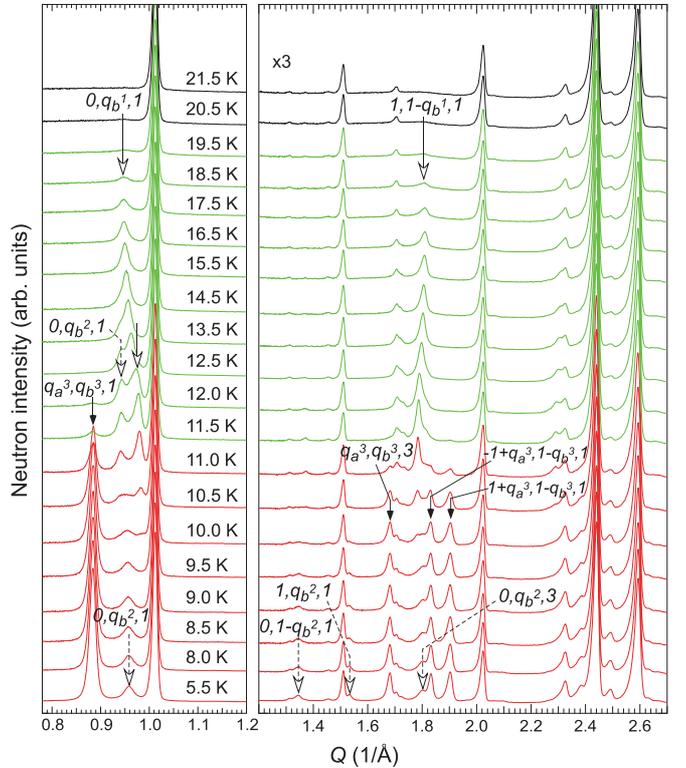

FIG. 3. (Color online) (a) Electric polarization and (b) relative dielectric constant of a powder sample of $2H$-AgFeO$_2$ as a function of temperature. The inset in (a) shows the poling electric field dependence of the polarization.

FIG. 4. (Color online) Temperature dependence of the neutron diffraction patterns for $2H$-AgFeO$_2$ collected on the WISH diffractometer. Intensity data in the higher range (right panel) are scaled by a factor of 3 to improve visibility.

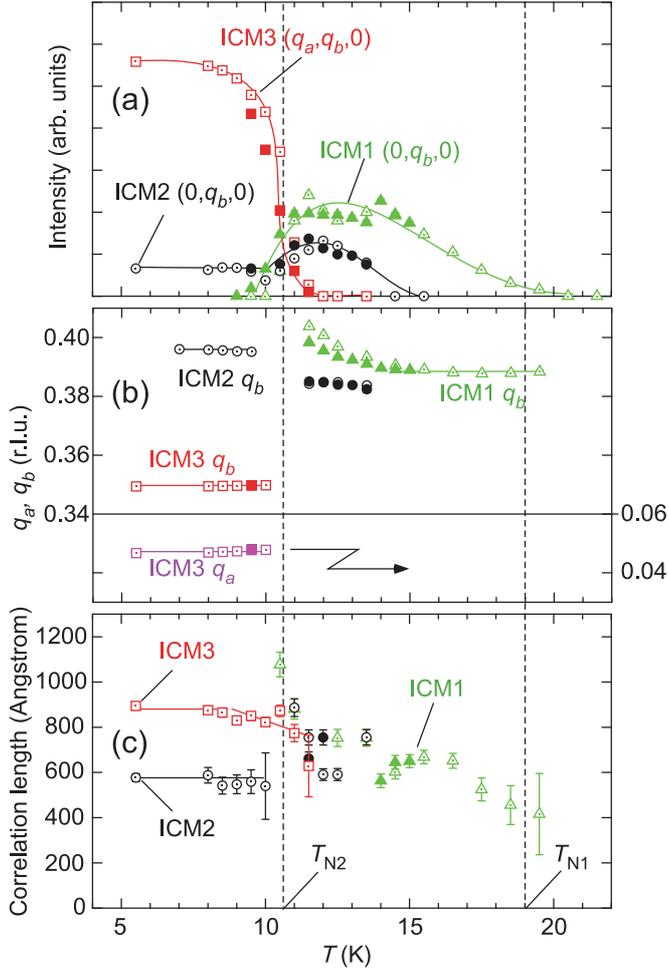

FIG. 5. (Color online) Temperature dependence of (a) integrated intensities of magnetic diffraction peaks, (b) propagation wave numbers, (c) correlation lengths of the magnetic orderings in $2H$-AgFeO$_2$. Open and closed symbols denote the data measured with heating and cooling processes, respectively.

Below $T \simeq 14$ K, an additional set of reflections was observed, as is clearly seen in Figs. 4 and 5(a). The reflections can be indexed with the $\bm{k}_2 = (0, q_b^2, 0)$ propagation vector and they survive down to 5.5 K (ICM2). The wave number $q_b^2$ sharply changes at $T_{N2}$, from $q_b^2 = 0.385$ ($T > T_{N2}$) to 0.395 ($T < T_{N2}$) [Fig. 5(b)]. The reciprocal lattice positions, where the magnetic satellite reflections appear for ICM2, are added to those for ICM1. While the magnetic satellites are only from the positions with $H + K = 2n$ in the ICM1 structure, e.g., $111 \pm \bm{k}_2$, the satellites in the ICM2 phase were observed at the positions from $H + K = 2n + 1$ as well as $H + K = 2n$, such as $011 \pm \bm{k}_2$ and $101 \pm \bm{k}_2$. This point will be discussed later. The correlation length for ICM2 is $\simeq 580$ Å [Fig. 5(c)].

The third set of magnetic reflections with an incommensurate wave vector appears below $T_{N2} = 11$ K (ICM3 phase), which is shown in Figs. 4 and 5(a). These reflections can be indexed by $\bm{k}_3 = (q_a^3, q_b^3, 0)$ with $q_a^3 = 0.047$ and $q_b^3 = 0.349$ ($\bm{k}_3 = [(q_b^3 - q_a^3)/2, (q_a^3 + q_b^3)/2, 0]$ in the hexagonal setting). The intensity for the ICM3 order is dominant at the lowest temperature compared to coexisting ICM2. We can thus conclude that the magnetic ground state of $2H$-AgFeO$_2$ is the ICM3 phase and the ferroelectric polarization below $T_{N2}$ is closely related to this magnetic order. Similar to $3R$-AgFeO$_2$ [21], the correlation length is finite in the hexagonal polytype and was estimated to be $\simeq 900$ Å at the base temperature 5.5 K. The observed finite correlation length might be caused by the coexistence of the different magnetic phases in $2H$-AgFeO$_2$.

### 2. Magnetic structure analysis

As shown in Fig. 6(a), we successfully refined the magnetic part of the diffraction data (collected with the 90° detector banks) at 14.5 K in the ICM1 phase of $2H$-AgFeO$_2$ using the collinear SDW model with spin moments along the $c$ axis, illustrated in Fig. 7(a). The parameter refined at 14.5 K is the amplitude of SDW, $M_{ac} = 3.12(3)$ $\mu_B$. Although we tried to refine the canting angle of the collinear moments from the $c$ axis, any improvements in the reliability factor were not obtained. The magnetic order parameter for the SDW structure is expressed by a time-odd irreducible representation of the $Ccmm$ space group, $mSM_3$ (in the ISODISTORT notations [28]) with the resultant $3+1$ magnetic superspace group $Amam1'(0,0,\gamma)s00s$, $\gamma = q_b^1$. Thus, the magnetic point group in the ICM1 phases is the centrosymmetric $mmm1'$, in agreement with the lack of polarization in the dielectric measurements [Fig. 3(a)]. The spin direction of the SDW structure is independent of temperature with the experimental accuracy, and different from that in the SDW structure with canting moments in $3R$-AgFeO$_2$ [21].

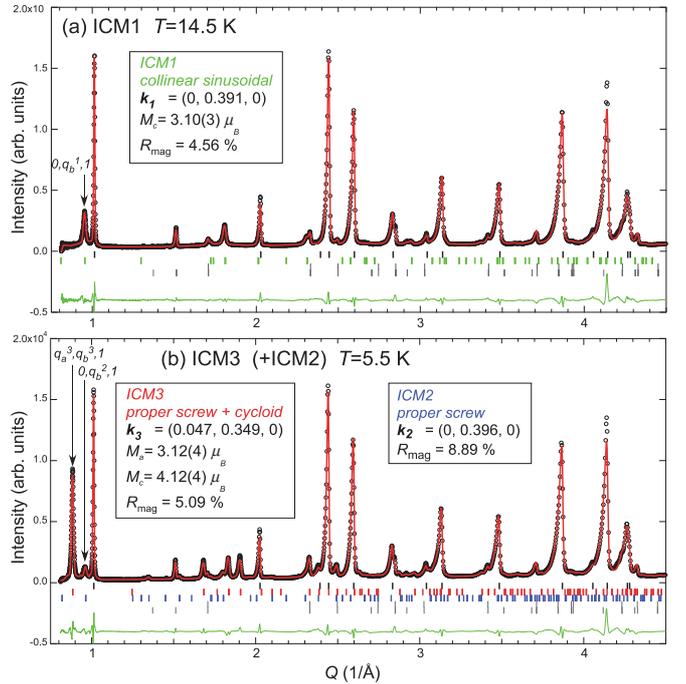

FIG. 6. (Color online) Typical result of the magnetic structural refinement for the experimental data at (a) 14.5 and (b) 5.5 K. The first line of the vertical bars indicate the peak positions for the crystallographic phase of $2H$-AgFeO$_2$. The second lines in (a) and (b), and the third lines in (b) correspond to the ICM1, ICM3, and ICM2 phases, respectively. The third and fourth in (a), and the fourth and fifth lines in (b) denote impurity phases of $\alpha$-Fe$_2$O$_3$ and the magnetic phase of $\alpha$-Fe$_2$O$_3$, respectively. The refined parameters and reliability factors are given in the insets.

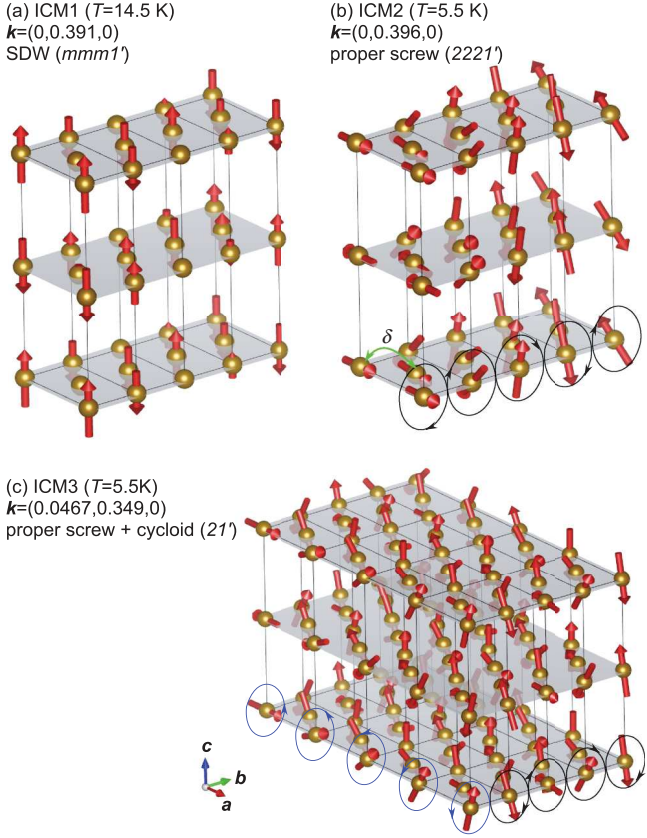

FIG. 7. (Color online) Illustration of the magnetic structures determined by the neutron diffraction measurements for $2H$-AgFeO$_2$. (a) Collinear spin-density-wave in the ICM1 phase, (b) proper screw in the ICM2 phase, and (c) spiral ordering in the ICM3 phase with the proper screw and cycloid components along the orthorhombic $b$ and $a$ axes, respectively.

The diffraction pattern of $2H$-AgFeO$_2$ measured at the lowest temperature 5.5 K was successfully refined by a mixture of proper screw structure [Fig. 7(b)] for the ICM2 phase and a combination of proper screw and cycloidal components [Fig. 7(c)] for the ICM3 phase. In the ICM2 phase, the satellite reflections for $k_2 = (0, q_b^2, 0)$ are observed at the positions from the reciprocal lattice points not only with $H + K = 2n$ (allowed by the $Ccmm$ symmetry), but also $H + K = 2n + 1$ (forbidden in $Ccmm$). This observation implies a presence of structural distortions violating the $C$-centering condition and indicates a further symmetry reduction at least down to $Pmma$. In this case, the phase difference in the magnetic modulation between the Fe sites, $(0,0,0)$ and $(\frac{1}{2}, \frac{1}{2}, 0)$ [$(0, 0, \frac{1}{2})$ and $(\frac{1}{2}, \frac{1}{2}, \frac{1}{2})$], are independent of each other. The phase difference was refined to be $\delta = 0.357 \pm 0.01\ \pi$, which is close to $\sim q_b^2 \pi$ corresponding to the ferromagnetic arrangement. The symmetry of the ICM2 phase is described by the $P2_1221'(0,0,\gamma)00ss$, $\gamma = q_b^2$ magnetic superspace group [28], indicating that the magnetic order parameter breaks all the mirror plane symmetries but keeps the two-fold rotational symmetries along the three orthogonal directions, resulting in the nonpolar magnetic point group $2221'$. This is consistent with the lack of the electric polarization in the temperature range of $11\text{ K} \leqslant T \leqslant 14\text{ K}$. In the refinement procedure, we constrained the values of the magnetic moments to be identical in the ICM2 and ICM3 phases. The magnetic structure of the ICM2 phase determined in the present work for the $2H$-AgFeO$_2$ polytype is similar to the magnetic polar phase found in CuFe$_{1-x}$Ga$_x$O$_2$ [15,20,22]. The magnetic point symmetries, however, are different, nonpolar $2221'$ in the former and polar $21'$ in the later cases. The difference relates to the hexagonal and rhombohedral parent symmetries, resulting in the absence and emergence of the ferroelectricity, respectively.

For the ICM3 phase, the determined magnetic structure involves both proper screw and cycloid orders, which is illustrated in Fig. 7(c). The $k_3$ propagation vector has two incommensurate components along the $a$ and $b$ directions, $(q_a^3, q_b^3, 0)$, and the spins are confined to be within the $ac$ plane. Thus, the proper screw modulation propagates along the $b$ axis, while the cycloid does along the $a$ axis. The spin components along the $a$ and $c$ axes in the spiral structure were refined to be 3.12(4) $\mu_B$ and 4.12(4) $\mu_B$, respectively, which indicates that the spiral has a small ellipticity. Unlike the proper screw ordering in ICM2, no violation of the $C$-centering condition has been observed for the ICM3 phase and the satellite reflections were found only at the allowed $H + K = 2n$ positions. The magnetic order parameter for the ICM3 structure combines two irreducible representations, $mP_1 \oplus mP_2$, of $Ccmm$, resulting in the $P2_11'(\alpha,\beta,0)0s$, $\alpha = (q_a^3 + q_b^3)/2$, $\beta = (q_a^3 - q_b^3)/2$ magnetic superspace group [28]. The corresponding magnetic point group is $21'$, which allows the ferroelectric polarization along the $c$ axis (this axis is identical in both orthorhombic and hexagonal cells).

## IV. DISCUSSION

Let us discuss the presence or absence of a ferroelectric polarization in $2H$-AgFeO$_2$ and $3R$-$A$FeO$_2$ polytyps. In the well-known spin current [6] and inverse DM [4,5] theories, the polarization is expressed by $\boldsymbol{p}_1 \propto \boldsymbol{r}_{ij} \times (\boldsymbol{S}_i \times \boldsymbol{S}_j)$, and is expected to be perpendicular to both $\boldsymbol{r}_{ij}$ and $\boldsymbol{S}_i \times \boldsymbol{S}_j$. In addition to $\boldsymbol{p}_1$, Kaplan and Mahanti have yielded another term, $\boldsymbol{p}_2 \propto \boldsymbol{S}_i \times \boldsymbol{S}_j$, which contributes to macroscopic polarization in both cycloidal and proper screw cases, unless a mirror plane containing $\boldsymbol{r}_{ij}$ or two-fold rotation axis perpendicular to $\boldsymbol{r}_{ij}$ exists [13]. In the case of $3R$-$A$FeO$_2$ ($A$ = Cu and Ag) delafossites with the $R\bar{3}m$ parent symmetry, the $\boldsymbol{p}_2$ term is allowed. The magnetic propagation vector $(q,q,0)$, specified in the hexagonal basis, breaks the three-fold rotational symmetry, resulting in the monoclinic $C2/m$ extended wave vector group. The further symmetry reduction depends on details of the magnetic order. In the $CuFe_{1-x}Ga_xO_2$ system, the magnetic structure is a proper screw type which keeps the twofold axis (magnetic point group $21'$ [22]) and induces the $p_2$ ferroelectric polarization parallel to $S_i \times S_j$ [15,20]. In $3R$-$AgFeO_2$, the magnetic structure is a cycloid which keeps the mirror plane (magnetic point group $m1'$) and allows both $p_2$ and $p_1$ components to be nonzero [21].

On the other hand, in $2H$-$AgFeO_2$ polytype, the parent space group is $P6_3/mmc$ and the extended wave vector group of the $(q,q,0)$ line of symmetry is orthorhombic $Ccmm$. This space group has three two-fold rotation axes, three mirror reflections parallel, and perpendicular to the orthogonal axes, respectively. In this case, the second term $p_2$ is not allowed, and the proper screw magnetic order has the nonpolar $2221'$ point group in the ICM2 phase, which yields neither $p_1$ nor $p_2$.

In the ICM3 phase, the spiral structure with proper screw and cycloid components results in the $21'$ magnetic point group and induces the polarization along the $c$ axis, which can be explained by the $p_1$ term. Consequently, although the crystal structures in the rhombohedral and hexagonal $ABO_2$ polytypes are similar, there is a significant difference in the mechanism generating the ferroelectric polarization based on the inverse DM effect.

We should also point out the huge difference in the values of the ferroelectric polarization between the ground states of the $2H$-$AgFeO_2$ and $3R$-$AgFeO_2$ polytypes. Whereas the polarization has been found to be $P \simeq 5$ $\mu C/m^2$ in $2H$-$AgFeO_2$, $3R$-$AgFeO_2$ shows two orders of magnitude higher $P \simeq 300$ $\mu C/m^2$ [21]. Comparing the $p_1$ component in the ferroelectric phases of $2H$-$AgFeO_2$ and $3R$-$AgFeO_2$, one can see a difference in the neighboring bonds generating the macroscopic polarization $P_1 = \sum_{ij}^N p_1^{ij}$. $N$ is the total number of bonds. Figure 8 illustrates a structural fragment in both phases, consisting of four spins from the $ab$ plane. In the case of cycloidal structure with the commensurate $a$ component $k = (-\frac{1}{2}, q, \frac{1}{2})$ ((ICM2) phase in $3R$-$AgFeO_2$ [21]), the local polarization generated by the two spins, $S_1$ and $S_2$ in Fig. 8(a), $p_1^{12} = \alpha \ r_{12} \times (S_1 \times S_2)$ is equal to $-p_1^{24}$ because $S_2 \times S_1 = -S_2 \times S_4$ ($\alpha$ is a constant). In the same way $p_1^{13} = -p_1^{43}$. Therefore, the local polarization on a bond parallel to $b$ axis, $p_1^{23}$, only contributes to the total polarization in $3R$-$AgFeO_2$. The contribution can be calculated as $P_1^{3R} = (N/3)\alpha S_2 S_3 \sin(2\pi q) \simeq 0.32 N\alpha S^2$. On the other hand, the spiral structure in the ICM3 phase of $2H$-$AgFeO_2$ has spins in the $ac$ plane and incommensurate wave vector component along both the $a$ and $b$ axes. Since the spins are confined within the $ac$ plane, any bond along the $b$ axis, e.g., $p_1^{23}$ in Fig. 8(b), does not generate a polarization. However, $p_1^{21}$ ($p_1^{13}$) is not canceled by $p_1^{24}$ ($p_1^{43}$) because $S_2 \times S_1 \neq -S_2 \times S_4$ ($S_1 \times S_3 \neq -S_4 \times S_3$), generating a net polarization along the $c$ axis. One can calculate that the macroscopic polarization is $P_1^{2H} = (N/3)\alpha S_2 S_1 \sin[\pi(q_b - q_a)] - (N/3)\alpha S_2 S_4 \sin[\pi(q_b + q_a)] \simeq 0.05 N\alpha S^2$ in $2H$-$AgFeO_2$. The calculated $P_1$ in the two polytypes could qualitatively explain the tendency $P^{3R} \gg P^{2H}$, but the estimated ratio, $P_1^{3R}/P_1^{2H} \simeq 6.4$ is not perfectly consistent with the experimental value $P^{3R}/P^{2H} \simeq 60$. The $p_2$ component that is allowed only in $3R$-$AgFeO_2$ might give the additional contribution to $P^{3R}$. For further understanding of the microscopic mechanisms generating the polarization in the different polytypes, more sophisticated theoretical calculations and single-crystal experiments are desirable.

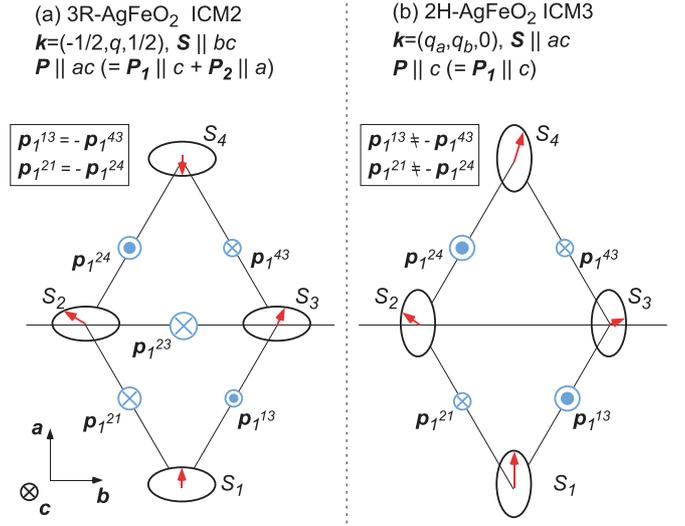

FIG. 8. (Color online) Schematic drawings of the spin arrangements and the generated $p_1$ components of the local polarization for each neighboring bond, derived from $p_1 = \alpha r_{i,j} \times (S_i \times S_j)$ in (a) $3R$-$AgFeO_2$ and (b) $2H$-$AgFeO_2$. The symbols representing the local polarizations roughly show their absolute values as a size of the circle.

## V. CONCLUSION

We have studied the magnetic and dielectric properties of the hexagonal triangular lattice antiferromagnet $2H$-$AgFeO_2$ by neutron powder diffraction, magnetic susceptibility, specific heat, pyroelectric current, and dielectric constant measurements. The successive magnetic phase transitions at $T_{N1} = 18$ K and $T_{N2} = 11$ K have been observed, which include the collinear (ICM1) and two noncollinear magnetic phases (ICM2 and ICM3). The ferroelectric polarization $P \simeq 5$ $\mu C/m^2$ emerges below $T_{N2}$ and it is two orders of magnitude smaller than that found in the rhombohedral $3R$-$AgFeO_2$ counterpart. In the ICM1 phase, stable in the temperature range of 11 K $\leqslant T \leqslant$ 18 K, the collinear SDW magnetic ordering with $k_1 = (0, q_b^1, 0; q_b^1 = 0.390$–$0.405)$ and magnetic moments along the $c$ axis is realized. The magnetic point group of the SDW ordering is $mmm1'$. The spin direction of SDW in $2H$-$AgFeO_2$ is different from that in $3R$-$AgFeO_2$, implying that the difference in the stacking sequence of the triangular layers significantly affects the magnetic anisotropy. Below 14 K, the incommensurate magnetic ordering of the ICM2 phase with $k_2 = (0, q_b^2, 0; q_b^2 = 0.385$–$0.396)$ appears and coexists with ICM1 or ICM3 as a minor phase. The determined magnetic structure of the ICM2 phase is the nonpolar proper screw with the magnetic point group $2221'$. Below 11 K, the ground-state spiral order with $k_3 = (q_a^3, q_b^3, 0; q_a^3 = 0.0467, q_b^3 = 0.349)$ appears and involves both cycloid and proper screw components. The magnetic point group is the polar $21'$ in agreement with the emergence of ferroelectric polarization observed in the pyroelectric current measurements. The ferroelectric polarization was concluded to be along the $c$ axis and induced by the inverse DM mechanisms through the $\bm{p}_1 \propto \bm{r}_{ij} \times (\bm{S}_i \times \bm{S}_j)$ relation [4–6]. Unlike $3R$-AgFeO$_2$, the additional $\bm{p}_2 \propto \bm{S}_i \times \bm{S}_j$ contribution to the macroscopic polarization does not exist in $2H$-AgFeO$_2$ due to the symmetry constraints imposed by the hexagonal lattice [13].

## ACKNOWLEDGMENTS

The images shown in Figs. 1 and 7 were depicted using the software VESTA [29] developed by Momma. The work at ISIS was supported by TUMOCS project. This project has received funding from the European Union's Horizon 2020 research and innovation programme under the Marie Skłodowska-Curie grant agreement No. 645660.